%% file: coel.tex
\documentclass[]{article}
\usepackage{authblk}
\pdfoutput=1

\usepackage{subfigure, graphicx, amsmath, amssymb} 
\usepackage{booktabs}
\usepackage{longtable}
\usepackage{hyperref}

\graphicspath{{images/}}

\begin{document}

\title{COEL: A Web-based Chemistry Simulation Framework}

\author[1]{Peter Banda\thanks{banda@pdx.edu}}
\author[2]{Drew Blount\thanks{dblount@reed.edu}}
\author[3]{Christof Teuscher\thanks{teuscher@pdx.edu}}
\affil[1]{Department of Computer Science, Portland State University}
\affil[2]{Artificial Life Lab, Reed College}
\affil[3]{Department of Electrical and Computer Engineering, Portland State University}

\maketitle

\begin{abstract}
The chemical reaction network (CRN) is a widely used formalism to describe macroscopic behavior of chemical systems. Available tools for CRN modelling and simulation require local access, installation, and often involve local file storage, which is susceptible to loss, lacks searchable structure, and does not support concurrency. Furthermore, simulations are often single-threaded, and user interfaces are non-trivial to use. Therefore there are significant hurdles to conducting efficient and collaborative chemical research. 

In this paper, we introduce a new enterprise chemistry simulation framework, COEL, which addresses these issues. COEL is the first web-based framework of its kind. A visually pleasing and intuitive user interface, simulations that run on a large computational grid, reliable database storage, and transactional services make COEL ideal for collaborative research and education.

COEL's most prominent features include ODE-based simulations of chemical reaction networks and multicompartment reaction networks, with rich options for user interactions with those networks. COEL provides DNA-strand displacement transformations and visualization (and is to our knowledge the first CRN framework to do so), GA optimization of rate constants, expression validation, an application-wide plotting engine, and SBML/Octave/Matlab export.
We also present an overview of the underlying software and technologies employed and describe the main architectural decisions driving our development.
COEL is available at \href{http://coel-sim.org}{coel-sim.org} for selected research teams only. We plan to provide a part of COEL's functionality to the general public in the near future.
\end{abstract}


\input{1_introduction.tex}

\input{1a_related_work.tex}
\input{2_functionality.tex}
\input{3_architecture.tex}
\input{4_conclusion.tex}

\section*{Acknowledgment}
This material is based upon work supported by the National Science Foundation under grant no. 1028120. We acknowledge Avi Debnath for his work on DNA-strand visualization, and GridGain for an academic license to use the GridGain HPC technology.

\vspace{40pt}

\bibliography{bibliography}
\bibliographystyle{splncs}

\end{document}

%% file: 1_introduction.tex
\section{Introduction}
\label{sec:introduction}

The main motivation behind the development of the COEL framework is the often monotonous and low-level management of scientific models. Further, running simulations on multiple threads and CPUs requires non-trivial effort. Research avenues built on solid theoretical ideas often run into trouble because of a lack of appropriate tools and software, leading to unnecessary delays, implementation of proprietary (home-made) solutions for basic tasks and reinventions of standard design patterns. As is true with most desktop applications, most existing tools provide access to only a single user on a local machine, requiring version-management software to enable collaboration, and general usability and visual appeal are usually low priorities. We argue that the way we work and conduct research must dramatically change to keep pace with the amount of data produced by simulations, to provide immediate and integrated visualization, and to enable geographically dispersed teams to work together on a single platform.

In this paper we introduce the \textit{COllective cELlular computing} (COEL) framework, the first web-based simulation framework for modeling and simulating chemical reaction networks (CRNs). COEL's web client is immediately accessible without any installation or download. The computational load of simulations is handled by COEL's grid rather than the client's machine. Remote teams can share and manipulate chemical models in real time. Data is stored remotely and safely in COEL's database, which is backed up daily. In developing COEL we emphasized platform-wide visualization, providing quick and embedded insight for users. 

It is important to emphasize the significance of COEL's database storage. Even though raw file storage (as opposed to structured databases) has been obsolete in industry for more than two decades, the scientific community still widely practices this approach. Storing data in files is not only ineffective, but its textual representation requires cumbersome parsing and tedious serialization for later structured searches or data mining. More so, files are inherently local, and without proper back-up, it is not uncommon that scientific data are lost. A recent study by Vines et al. in Current Biology~\cite{Vines2014} found that $80\%$ of scientific data are lost within two decades, disappearing into old email addresses and obsolete storage devices. Alarmingly, the authors found that the average rate of data loss is $17\%$ each year.
Furthermore, because of private and local storing only $11\%$ of the academic research in the literature was reproducible by the original research groups, as reported in Nature~\cite{Begley2012}. This is intuitively more prevalent in experimental science, but computer-based research is affected as well.
We suggest that with current scientific approaches this problem will only worsen in the age of big data. We argue that storing all (even intermediate) models and results remotely and in a reliable long-term fashion, and making them accessible to the general scientific community should become the new standard. With remote data storage and a convenient web client, users do not have to deal with version-compatibility of data structures, as it is the case with traditional approaches. Since a new application release is deployed together with a central migration of the database, version updates are worry-free for users. 

Accessibility has two important consequences: collaboration and transparency. Using COEL, as with so-called `cloud-based' web applications, individuals can work on different facets of the same project and see each other's modifications in real-time. This has allowed the authors of this paper, for example, to study the same system, run parameter evolutions and performance evaluations, modify simulation dynamics and so on from separate campuses.

COEL has been developed as a part of the NSF project ``Computing with Biomole-cules". We have successfully applied COEL as a sole tool to model and evaluate various types of chemical perceptrons~\cite{Banda2013,Banda2014,Banda2014b}, chemical delay lines and time-series learners~\cite{Banda2014c,Moles2014}, and random DNA circuits~\cite{Banda2013b}.

In this paper we first discuss the state-of-the-art in chemistry simulation frameworks (Section \ref{sec:related-work}), then present COEL's functionality (Section \ref{sec:functionality}) and technical architecture (Section \ref{sec:architecture}). We conclude with a discussion of COEL's place in the ecosystem of chemistry simulation frameworks, and the future of COEL (Section \ref{sec:conclusion}).

%% file: 1a_related_work.tex
\section{Related Work}
\label{sec:related-work}

COEL is not the first software made to simulate chemical reaction networks. There are already many programs which do so, and together the field of CRN simulators \cite{Hoops2006,castellini2009,Schmidt2006,Funahashi2008,Hucka2003} offers a huge set of technical features, e.g., simulation options and statistical tools. Our goal with COEL was not (so much) to introduce new simulation algorithms or methods of analysis, but to include the most common and useful tools among CRN simulators in an intuitive and modern web-based package. This makes the tools of systems biology more accessible, and the research done with them more transparent, collaborative, and replicable.

COPASI~\cite{Hoops2006} is arguably the most advanced and widely used tool. In a nutshell, COPASI simulates a variety of chemical objects and allows for freedom in experiment design and statistical analysis. COPASI is quite feature rich, and could be considered the gold standard of CRN simulation frameworks. There are others worth mentioning, of course, such as those in the MATLAB Systems Biology Toolbox~\cite{Schmidt2006}, and CellDesigner~\cite{Funahashi2008}, which is a modeling tool for biochemical networks. Most of these tools share support for the SBML language for describing chemical systems~\cite{Hucka2003}, which as a standard has been a great boon to the field, enabling cross-platform migration.

Along with SBML support, most simulation environments share a core set of capabilities. Beyond basic deterministic ODE integration of CRNs (and stochastic reactions, a feature which COEL notably does not have), it is common to offer parameter optimization to help in the design of the networks themselves. Programs such as COPASI and CellDesigner can simulate a number of other biochemical objects of interest, such as cellular compartments. It is common to allow for various kinetic models of chemical interactions, such as Michaelis-Menten \cite{michaelis13} and mass action \cite{lotka1920}.

In many kinds of frameworks, there is some tension between the depth of features and the features' accessibility, especially for highly technical applications such as CRN simulators. In addition to offering rich design capabilities, many developers of CRN simulators have the explicit motivation of reaching a large audience: The authors of COPASI said, ``... the software needs to be available for the majority of scientists ..." (p. 3069,~\cite{Hoops2006}). The authors of CellDesigner felt similarly, saying that they wish to "confer benefits to as many users as possible" (p. 1255,~\cite{Funahashi2008}). COEL automatically runs on any operating system with a web browser, including smartphones or tablets, so it is accessible anywhere in the world without any installation. Further, COEL's computational grid centrally runs any difficult tasks which might run slowly on clients' computers. We strongly believe that there is no more accessible paradigm for research tools than a web-based interface with computation performed in the cloud.

%% file: 2_functionality.tex
\section{Features and Functionality}
\label{sec:functionality}

COEL provides a unified web environment for the definition, manipulation, and simulation of chemical reaction networks. In this section, we will discuss COEL's functionality and application-wide features in detail.

\subsection{Chemical Reaction Network Definition}

At its most basic level, a chemical reaction network (CRN) consists of a finite set of chemicals and reactions. A CRN represents an unstructured macroscopic simulated chemistry, hence the species labeled with symbols are not assigned a molecular structure. The state of a CRN is represented by a vector of chemical species concentrations. 

Each reaction is of the form $a_1X_1 + \ldots + a_n X_n \rightarrow b_1 Y_1 + \ldots + b_m Y_m$, where species $X_i$ are reactants and $Y_i$ products. Constants $a_i$ and $b_i$ are stoichiometric factors, i.e., positive integers describing how many copies of each molecule are involved in the reaction. For instance the reaction $A + B \rightarrow C$ describes species $A$ and $B$ binding together to form species $C$. Reactions can also involve catalysts or inhibitors, which speed up or slow down the reaction, but are not consumed.

Note that a legal reaction could have no reactants or no products. For that purpose we include a special no-species symbol $\lambda$ to represent a formal annihilation $A + B \rightarrow \lambda$ or a decay $A \rightarrow \lambda$.  Mass conservation states that matter cannot be destroyed nor created, i.e., in a closed system the matter consumed and produced by each reaction is the same. Annihilation and decay as we defined them seem to violate that, however, in the chemical analogy, $\lambda$ does not signify a disappearance of matter but simply an inert species, effectively absent from the system of chemical interactions. Similarly we interpret a reaction $\lambda \rightarrow A$ as an influx of $A$ rather than a creation of a molecule $A$ from nothing.

Reaction rates define the strength or speed of reactions, as prescribed by kinetic laws--Michaelis-Menten~\cite{cop02} kinetics for catalytic reactions, and mass action kinetics~\cite{espenson95} otherwise. The rate of an ordinary reaction $a_1S_1 + a_2S_2 \rightarrow P$ is defined by the mass-action law as
\begin{equation*}
 r = \frac{d[P]}{dt} = -\frac{1}{a_1}\frac{d[S_1]}{dt} = -\frac{1}{a_2}\frac{d[S_2]}{dt} = k[S_1]^{a_1}[S_2]^{a_2},
\end{equation*} 
where $k \in \mathbb{R}^+$ is a reaction rate constant, $a_1$ and $a_2$ are stoichiometric constants, $[S_1]$ and $[S_2]$ are concentrations of reactants (substrates) $S_1$ and $S_2$, and $[P]$ is a concentration of product $P$. The rate of a catalytic reaction $S \xrightarrow{E} P$, where a substrate $S$ transforms to a product $P$ with a catalyst $E$, whose concentration increases the reaction rate, is given by Michaelis-Menten kinetics as
 \begin{equation*} 
r = \frac{d[P]}{dt} = \frac{k_{cat} [E][S]}{K_m + [S]},
\end{equation*}
where $k_{cat}, K_m \in \mathbb{R}^+$ are rate constants.

COEL is consistent with these general CRN formalisms; next, we will describe details particular to COEL's implementation. COEL automatically computes appropriate rate functions once given numeric rate constants, yet it also allows users to define arbitrary rate functions using custom expressions over species labels, giving the user full freedom over the system's dynamics. Reactions can be uni- or bidirectional, and bidirectional reactions can have independent forward and backward rates.

\begin{figure}[t!]
\centering
    \includegraphics[width=\textwidth]{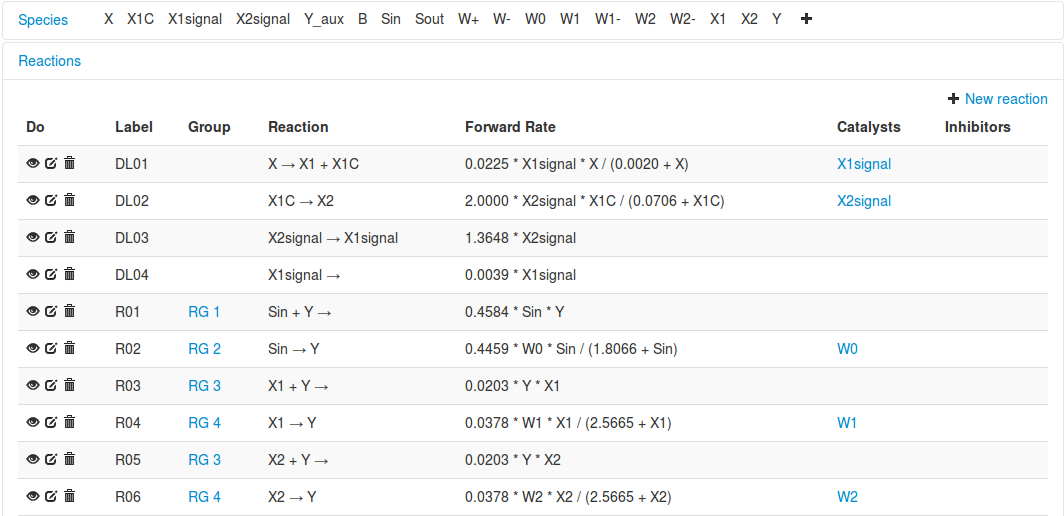}
\caption{A partial description of a chemical reaction network in COEL. Species are listed at the top, and their reactions  are presented in tabular form. The reactants and products are described in the third column, the forward reaction rates are in the fourth column, and any catalysts are in the fifth.}
\label{fig:reaction_set}
\end{figure}

Both species sets and reaction sets are extensible, in that new sets can be defined as expansions of old ones. This promotes reuse and modular design. Further, two CRNs can be merged combining their reactions and species into one network.

Figure \ref{fig:reaction_set} shows an example CRN in COEL, a memory-enabled chemical perceptron~\cite{Banda2014c}. The CRN's species, reactions, and reaction rates are presented in a unified view from which any of these objects can be easily edited in a few steps. Also, users can export CRNs in Matlab, Octave, or SMBL formats if they wish to study their systems using different tools. It is also possible to import an SBML-defined CRN into COEL.

\begin{figure}[t!]
\centering
    \includegraphics[width=0.55\textwidth]{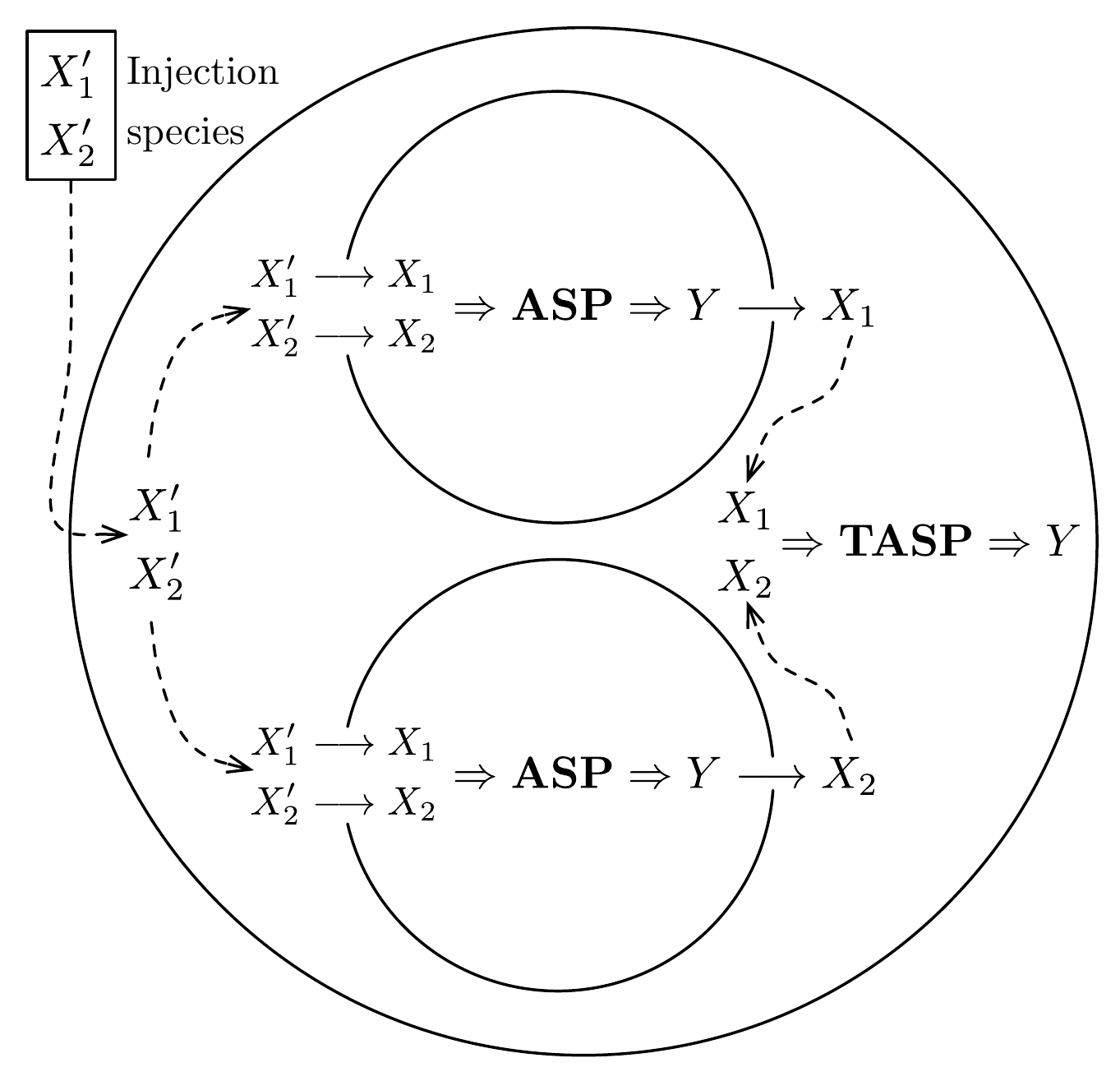}
\caption{Schematic of permeation in a simple 2-1 multicompartment system from one of the authors' current projects. The `tagged' input species $X_1'$ and $X_2'$ are injected into the outer compartment. They permeate into the inner compartments via channels which transform them into regular, untagged input species $X_1$ and $X_2$. The inner compartments' ASPs (Asymmetric Signal Perceptrons \cite{Banda2014}, each of which is a large CRN) process the input species into the output $Y$. Each compartment has a unique outgoing channel to transform $Y$ into one of the input species, which are then processed in the outer compartment.}
\label{fig:feedingforward}
\end{figure}

\begin{figure}[t!]
\centering
    \includegraphics[width=\textwidth]{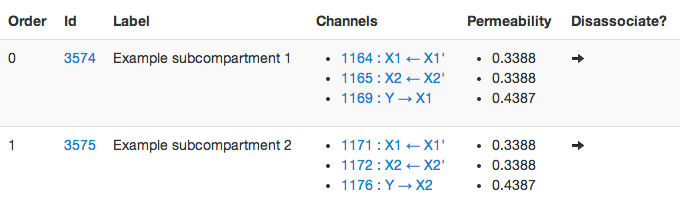}
\caption{COEL's representation of the permeation schema depicted in Figure \ref{fig:feedingforward}.}
\label{fig:compartments}
\end{figure}

In imitation of biochemical cells or membranes, CRNs in COEL support hierarchical tree-like compartmentalization. Each compartment hosts an independent reaction set and vector of chemical concentrations. Compartments communicate with each other through permeation, formalized in what we call `channels.' A channel works just like an ordinary reaction, except the reactant and product species reside in adjacent compartments. Among other things, this allows for modular design of chemical systems, where connected modules reside in nested compartments, as shown in Figures \ref{fig:feedingforward} and \ref{fig:compartments}.

\begin{figure}[t!]
\centering
    \includegraphics[width=\textwidth]{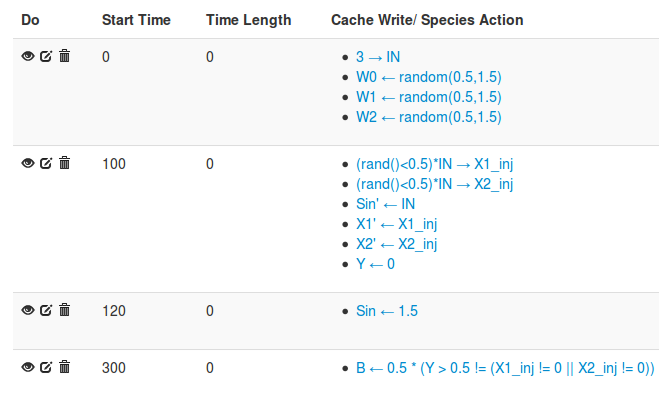}
\caption{The details of a COEL interaction series. Left arrows denote the setting of species concentrations, and right arrows indicate assignments of user-defined variables. The interaction at time 100 does the following (note that at time 0 the variable $IN$ is set to 3): first, the variables $X1_{inj}$ and $X2_{inj}$ are randomly set to 0 or 3 with equal probability. The concentration of $Sin'$ is set to 3, then the concentrations of $X1'$ and $X2'$ are set equal to their respective injection variables. Finally, $Y$ is flushed from the system---its concentration is set to 0.}
\label{fig:interaction-series-example}
\end{figure}

\subsection{Chemical Reaction Network Simulation and Interaction Series}

A major feature of COEL, in that it has been crucial to its early users and their work, is so-called interaction series. An interaction series allows the user to directly manipulate concentrations of species in the CRN. This feature is analogous to, though more capable than, automatic chemical injections into a reaction chamber. For compartment-extended CRNs, interaction series can be identically hierarchical, allowing for precise interaction with each component of the network.

Concentrations can be modified multiple times, not just initially. E.g., for iterative processes it is useful to define a set of periodic interactions. In specifying interactions, a user can define custom concentration-setting expressions, as well as custom variables for use in those expressions. For example, the bottommost interaction in Figure \ref{fig:interaction-series-example} injects species $B$ (here a `penalty species') at concentration $0.5$ if the output species $Y$ does not match {\small \ttfamily AND} of the original input concentrations, $X1_{inj}$ and $X2_{inj}$. The COEL Interaction Series API, as we call it, is then a scripted language that can describe a variety of complicated experimental scenarios without touching the underlying simulation-framework code. Thus end users have the freedom to manipulate the chemical system in a dynamic and safe way (basic expression validation is provided). 

\begin{figure}[htb!]
\centering
    \includegraphics[scale=.4]{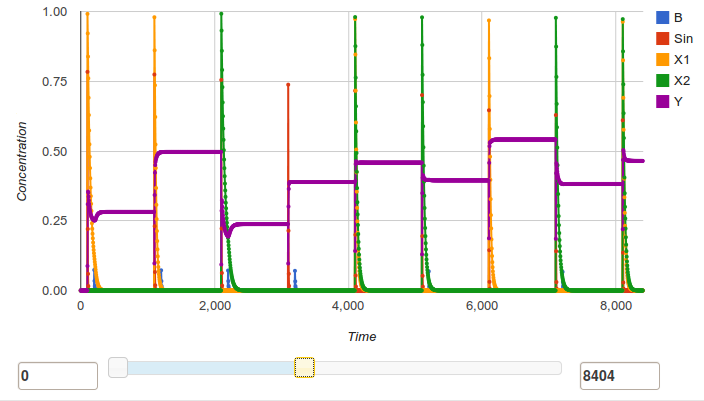}
\caption{A chart showing concentration traces of $5$ chemical species over time in COEL. In this case, an interaction series injects a random combination of $X1$ and $X2$ at concentration 1, every 1000 time steps.}
\label{fig:concentration-chart-example}
\end{figure}

To actually simulate a CRN, a user runs a defined reaction network with a selected interaction series (which might be as simple as setting initial concentrations). Users can choose from a number of non-adaptive and adaptive deterministic ODE solvers to integrate their system. Upon running such a simulation, the user is by default shown an embedded chart of species concentrations over time (Figure ~\ref{fig:concentration-chart-example}). If further post-processing is required, full or filtered data could be easily exported into a CSV file.

Note that since ODE solvers are deterministic, two simulations using the same CRN and interaction series will always produce the same concentration traces if the interaction series is deterministic. That is, however, not the case for the interaction series in Figure \ref{fig:interaction-series-example}, which uses random weight setting and randomly injects binary inputs at concentration $0$ or $3$. COEL does not currently have a feature to save random number seeds to exactly replicate simulations such as these.

\subsection{Performance Evaluation and Dynamics Analysis}

COEL provides a core set of tools for analyzing and modifying CRNs, enabling statistical record-keeping as well as the design of complex networks whose precise architecture is initially unknown to the user. COEL's basic interpretive tool is the ``translation series,'' defined by the user in a similar manner to interaction series, described above. A single translation is a straightforward function of the current concentrations and any predefined constants, and can be Boolean or numeric in its output.

One can simply plot the output of a translation series to see the CRN's behavior through a certain lens, or use the series as the basis of evaluation and optimization. Because many CRNs involve a random component, especially in (but not limited to) their interaction series, COEL allows the user to run large batches of simulations and collect statistics based on these translation series.

Because it is usually difficult to precisely translate simulated chemistries into wet ones, COEL also offers perturbation analysis. Users can evaluate the performance of the CRN if a defined set of rates are randomly perturbed according to set parameters. This is useful in measuring the robustness of a chemical system.

COEL also offers dynamics analyses with a detailed statistical view of an individual CRN simulation. This includes Lyapunov exponents, Derrida stability, time and spatial nonlinearity errors, and more; along with reports about the simulation itself, like how many species concentrations reached fixed points for given tolerance.

\begin{figure}[htb!]
\centering
    \includegraphics[width=\textwidth]{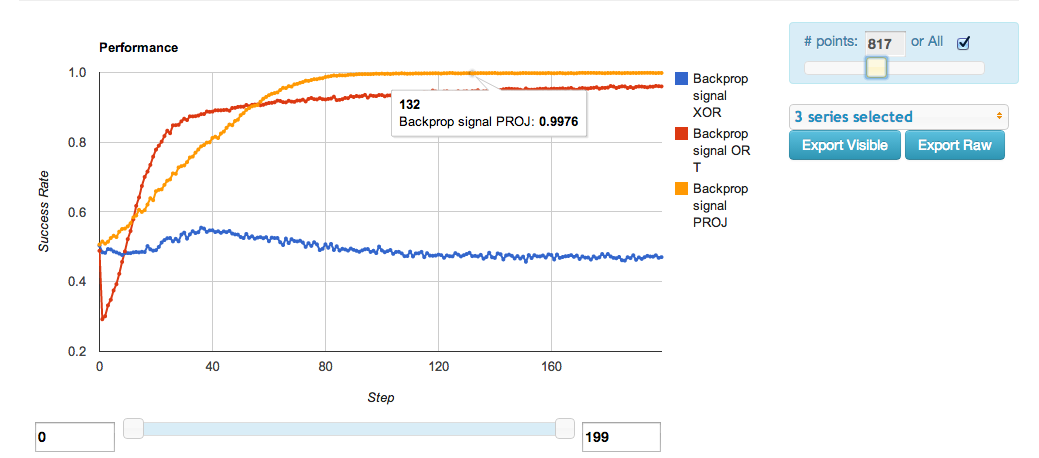}
\caption{A chart of three separate performance evaluations, each one showing the performance of a binary chemical perceptron averaged over 10,000 repetitions for given interaction series representing desired binary function ({\small \ttfamily XOR}, {\small \ttfamily OR}, {\small \ttfamily PROJ}). Note the data export options on the right.}
\label{fig:PerfEval}
\end{figure}

To allow maximum freedom in analysis, COEL offers CSV export of any raw data a user might produce. Every chart and data visualization in COEL is accompanied by a CSV export function, allowing the user to export either the data currently displayed on-screen (to replicate a chart or precisely modify its appearance) or the entire raw dataset, as shown in Figure \ref{fig:PerfEval}.

\subsection{Rate Constant Optimization}

With defined evaluation criteria, a user can optimize CRN's parameters with COEL's flexible genetic algorithm tool. Users define the space to be optimized by selecting which reaction and channel permeation rates are to be modified, in what ranges, and under what constraints (e.g. several reaction rates can be fixed to each other). Chromosomes are then vectors of rate constants.

The parameters of COEL's GAs are easily modified, allowing for different rates of mutation, rules of reproduction, initial populations, and so on. Chromosomes can be selected to reproduce either deterministically with elite selection, or probabilistically relative the measured fitness of each chromosome. Reproduction can be sexual or asexual. In the former case, crossover between two chromosomes can be either one-point (i.e., in chromosomes of length $n$, the child's first $p\leq n$ genes are from one parent and the last $n-p$ are from the other), or a probabilistic shuffle. Supported mutation types are one-bit, two-bit, exchange and per-bit, with content replacement and perturbation options. COEL's GAs also support fitness renormalization, and selection of maximization or minimization of the target function (fitness vs. error).

\begin{figure}[htb!]
\centering
    \includegraphics[width=\textwidth]{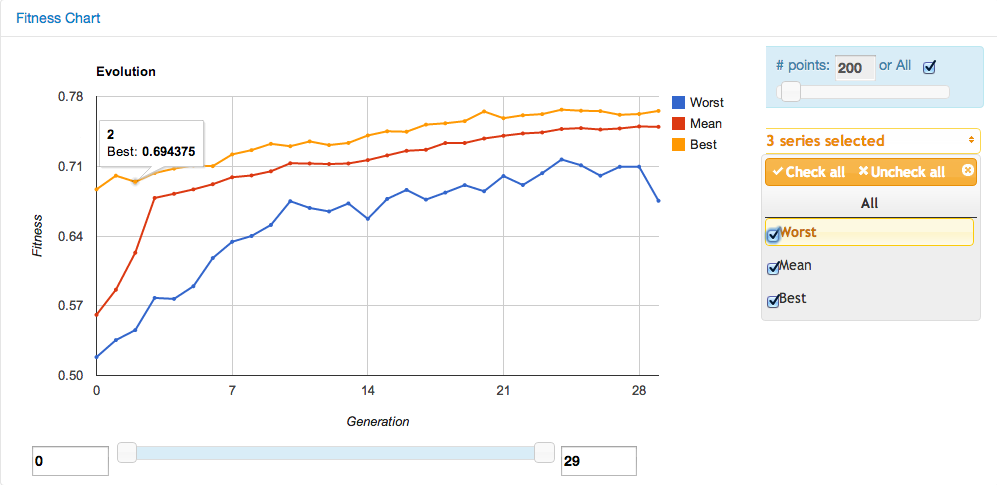}
\caption{A chart of a population's fitness over time in a run of a particular GA. This plot displays several features shared by all plots in COEL, enabling modification of the plot without refreshing the web page: an x-axis slider to specify the plot's domain, a drop-down menu to select which series to display, and a slider to select the plot's resolution relative the data set.}
\label{fig:evo}
\end{figure}


\subsection{DNA Strand Visualization and Displacement Reactions}
\label{subsec:dna}

COEL has a convenient web interface for visualizing DNA strands specified by the Microsoft Visual DSD syntax~\cite{Lakin2012b}, which decomposes single and (full or partial) double DNA strands into labeled subsequences called domains. Domains are classified as either long or short, also called toeholds. These DNA-strand images can be exported in the svg format, appropriate for publications and educational purposes alike. Note that the Microsoft Visual DSD web tool (unlike COEL) requires an installation of Microsoft Silverlight, whose support on Linux is problematic.

\begin{figure}[t!]
\centering
    \includegraphics[width=0.7\textwidth]{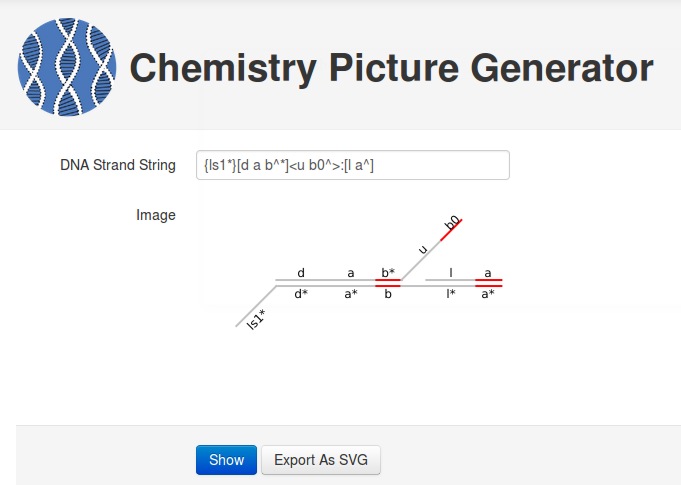}
\caption{COEL's tool for visualizing DNA strands specified in Visual DSD. Red lines represent toeholds, and gray lines are long domains.}
\label{fig:chem_pic}
\end{figure}

Furthermore, COEL can transform any CRN based on mass-action kinetics into a DNA strand-displacement circuit using the methods of Soloveichik et al. \cite{sol10}. In strand displacement systems, populations of these species are typically represented by the populations of single-stranded DNA molecules. These interact with double-stranded gate complexes which mediate transformations between free signals. In a nutshell, the mass-action reaction $X_1 + X_2 \rightarrow X_3$ is translated to three displacement reactions $X_1 + L \leftrightharpoons H + B$ (a single strand $X_1$ displaces an upper strand $B$ from the complex $L$), $X_2 + H \rightarrow O + W_1$ (a single strand $X_2$ displaces an upper strand $O$ from the complex $H$), and finally $O + T \rightarrow X_3 + W_2$ (a single strand $O$ displaces an upper strand $X_3$ from the complex $T$), where $L, H, B, O, T, H$ are auxiliary fuel species, and $W_1$ and $W_2$ are waste products.

\begin{figure}[htb!]
\centering
    \includegraphics[width=0.95\textwidth]{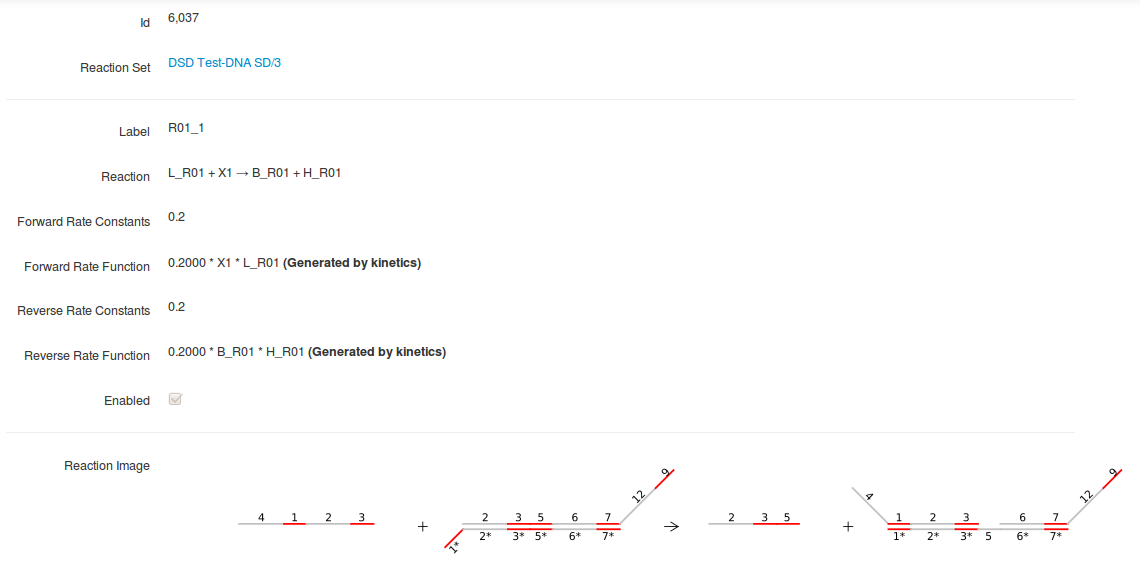}
\caption{A DNA strand displacement reaction obtained by COEL's transformation of arbitrary CRNs into strand displacement circuits.}
\label{fig:dna}
\end{figure}

Once applied to a reaction set, the transformation produces a CRN with new intermediate species and reactions, describing displacements of single strands from partial or full double strands. Besides new reactions, COEL also specifies the DNA structure of each species in terms of numerically-labeled domains, the output of which is shown in Figure \ref{fig:dna}. This is a powerful tool for automatic translation of so-called \textit{in silico} systems to feasible wet chemistries in a user-friendly way. The authors are not aware of any other CRN simulation framework that includes DNA strand displacement transformations as a part of their application toolbox.

\subsection{Random Chemical Reaction Network}

COEL offers functionality to quickly make a random chemical reaction network with set specifications. User-defined parameters include the number of species, the number of reactions, the number of reactants and products in each reactions, and a random distribution of reaction constants; COEL meets all of these constraints with combinatorial design. For open systems the user can also specify influx and efflux constraints.

Furthermore, COEL also supports generation of random DNA-stand circuits~\cite{Banda2013b} using single, full double, and partial double strands. Parameters for this function include number of single strands, ratio of upper to lower strands, ratio of upper strands with complements, (positive) normal distribution of partial double strands per upper strand, (positive) normal distribution of rate constants, ratio of influxes and effluxes, and distribution of rate constants. Based on a randomly generated ordering, DNA strands with higher order take precedence over lower-order strands in DNA-strand displacement reactions (Section~\ref{subsec:dna}). Also, note that the maximum number of strands that could bind together is two, which is justified by assuming that a single strand does not bind to partial double strand, but always displace its upper or lower part. We assume wet synthesis of these networks is possible by standard DNA sequence design~\cite{Zadeh2010}.

\subsection{Platform-wide Features}

Numerous features of COEL are omnipresent throughout the platform, creating a familiar look-and-feel as well as providing intuitive access to common features. Throughout COEL, users input mathematical functions in the straightforward syntax of the Java Expression Parser (displayed in Figure \ref{fig:interaction-series-example}), and those expressions are always validated by COEL before being input into any simulation. Views, such as COEL's list of reaction sets or interaction series, have a common search and filter feature, allowing for easy navigation through huge sets of objects. 

All charts in COEL are made with the Google Charts API, and include sliders for domain selection and data filtering (see Figure \ref{fig:evo}), as well as CSV export options (see Figure \ref{fig:PerfEval}). Finally, COEL has rudimentary user privacy protocols, where each user account is either a `user' who can see only his/her own projects, or an `admin' who can see every project on COEL. In order to share a project, a group of users currently have to have admin rights. We plan to expand privacy features in later versions.




%% file: 3_architecture.tex
\section{Architecture and Technology}
\label{sec:architecture}

COEL's architecture is highly modular with strict separation of business logic and technological application aspects. Nowadays, the main challenge of enterprise application development is not programming per se but rather the integration of diverse technologies and libraries which each address different application needs. The absence of strict inter-modular / inter-layer dependencies enables quick and easy customization and replacement of technologies and providers.

\begin{figure}[p!]
\centering
    \includegraphics[scale=.54]{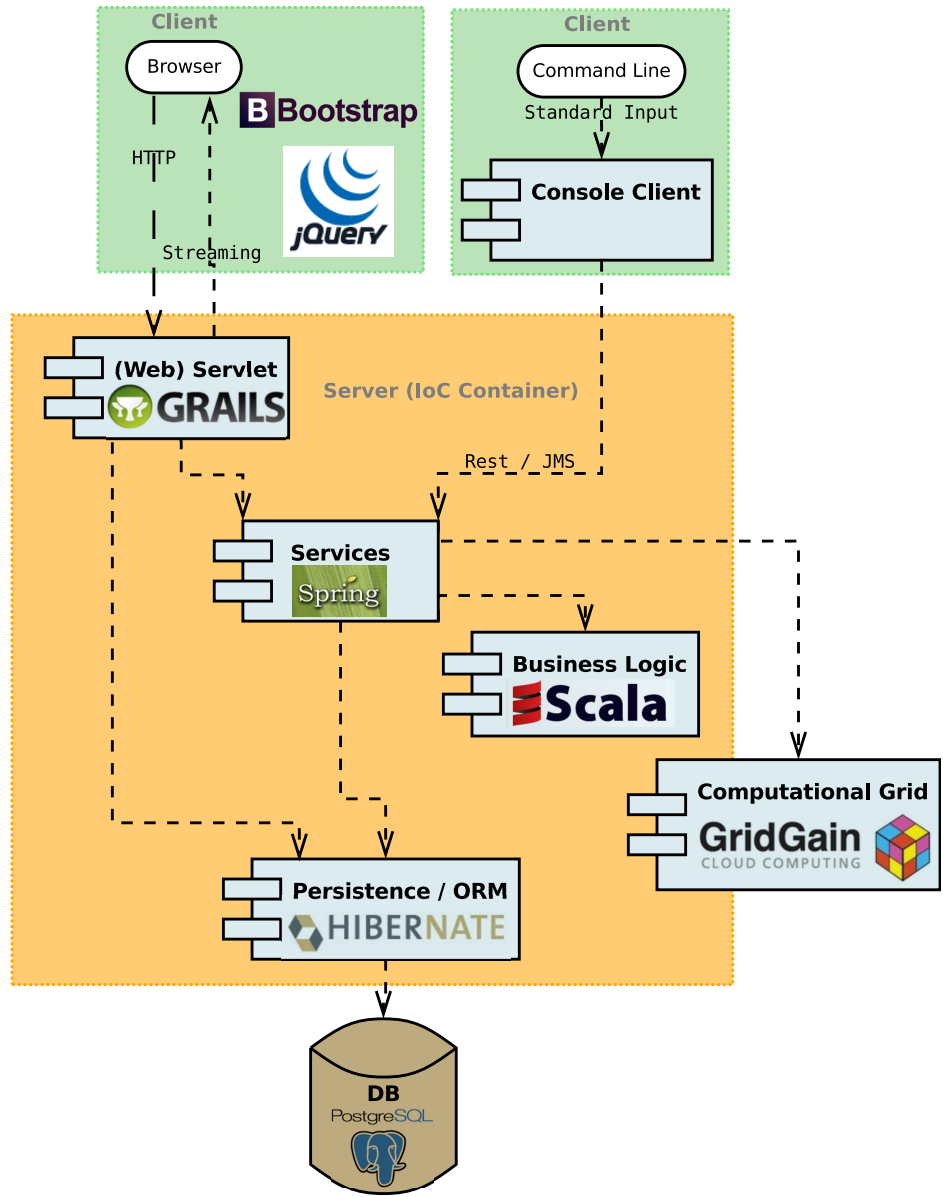}
\caption{A high-level overview of COEL's architecture consisting of web and console clients, web servlet, services, business logic, persistence layer, and computational grid. The application (IoC) container holding the server-side of the application is implemented in Spring framework.}
\label{fig:architecture}
\end{figure}

At this level of abstraction only the domain objects, the data holders of business data, implemented as POJOs (Plain Java Objects), are shared among all application parts and layers. Figure ~\ref{fig:architecture} presents a high-level overview of COEL's architecture with call (request) pathways. 
On the very top we have two clients representing the only entry points to the application: the web client backed by Grails \cite{grails}, jQuery \cite{jquery} and Bootstrap \cite{bootstrap} frameworks (discussed in Section \ref{subsec:grails}), and the plain console client implemented in standard Java for ``headless'' scripting.

Based on user's requests, the clients call the services such as {\small \ttfamily ChemistryService}, {\small \ttfamily EvolutionService}, and {\small \ttfamily UserManagementService} (Section \ref{subsec:service}) maintained by the Spring application container (Section \ref{subsec:ioc}), which then redirects either to a computational grid implemented on the top of GridGain HPC technology \cite{gridgain} (Section \ref{subsec:cloud_computing}) for distributed task execution, or to the persistence layer with DAOs (Data-Access Objects) and ORM (Object-Relation Mapping) provided by Hibernate \cite{hibernate} (Section \ref{subsec:persistence}). In addition, the web client controllers have a direct link to the persistence layer, which is beneficial especially for basic CRUD (Create, Read, Update, Delete) operations. At the very bottom a PostgreSQL \cite{postgresql} database stores and provides data on the demand of the persistence layer.

The business logic such as chemistry simulation and GA optimization is implemented mainly in the Scala language, leveraging both object-oriented and functional programming approaches. All technologies and libraries integrated into COEL are either open-source or free to use.

\begin{table}[htb!]
  \centering
  \caption{A list of the acronyms used in this section.}
  \begin{tabular}{l|r}
    \toprule
    \multicolumn{1}{c|}{Acronym} & \multicolumn{1}{l}{Description} \\
	\hline
JVM               & Java Virtual Machine \\

ORM               & Object-Relational Mapping \\

POJO              & Plain Java Object \\

DAO               & Data-Access Object \\

IoC               & Inversion of Control \\

JEP               & Java Expression Parser \\

JMS               & Java Message Service \\

REST              & Representational State Transfer\\

HPC               & High Performance Computing\\

JDBC              & Java Database Connectivity \\

SQL               & Structured Query Language \\

PLSQL             & Procedural Language/Structured Query Language \\

HQL               & Hibernate Query Language \\
  \bottomrule
  \end{tabular}
  \label{tab:acronym}
\end{table}

\subsection{Application Container}
\label{subsec:ioc}


The Spring Framework~\cite{spring,walls2010} provides the COEL's core application infrastructure. Spring is a leading enterprise solution for Java maintained by the SpringSource community since 2002. Compared to Enterprise Java Beans, the Spring portfolio is less invasive and more flexible. Spring is not an application server, it is just a set of libraries which can be used and deployed anywhere (like e.g., Tomcat and Jetty). It consists of several sub-projects which can be used separately or together as needed. Spring is a lightweight tool that shows how little is really needed for enterprise application development. It does not have strict dependencies, and it detaches technical and business concerns.

The IoC (Inversion of Control) container is a central part of the Spring Framework. It controls the creation, number of instances (with singleton and prototype scopes), lifecycle, inter-dependencies (loose-coupling or wiring) and general configuration of application components, modules, adapters, specific utility classes or in general any POJO whose creation and use should be maintained in the application context. Spring IoC is a simple and transparent glue or integrator of various components and frameworks which are provided either by Spring Portfolio itself or other parties. 

The IoC container encourages the best practices of programming with interfaces, i.e., each bean (POJO object in the IoC container) should consist of an interface and implementation class. Therefore, each bean knows that it can talk to a different bean that does something specific, but not which type of object, how its functionality is implemented, nor how the call is carried out. The IoC container injects the dependencies into POJOs at the runtime, and so beans take care only about their business purpose, not creation (and maintenance) of their relationships. 

This approach is superior to the factory design pattern because all dependencies get injected and configured through the application container (annotations and/or XML), however beans are not aware of the container's existence, i.e., unlike the factory pattern they do not need to call the application container in order to get their dependencies. The application code of Spring beans has little dependency on Spring itself. As a matter of fact, IoC is often described with the Hollywood principle: ``Don't call us, we call you." Besides Spring, other popular IoC containers include GUICE and Pico.

IoC abstraction results in modular, lightweight and layered architecture with loose-coupled pluggable components. Programmers are also encouraged to implement beans as thread-safe and stateless if possible, so several callers could safely query the same component without worrying about timing and/or call history. 

Last but not least, Spring IoC enables COEL to become a truly test-driven project. Because of loose-coupling and dependency injections, our JUnit tests could switch to test (rather than production) application context and substitute for instance implementation classes that require remote access to production systems with mock objects.


\subsection{Services}
\label{subsec:service}

The service layer is the actual gateway to the business/functional part of the application. Services are callable functions provided to the clients (or outside world). COEL is divided into five functional modules, each exposed by a separate service interface (facade): {\small \ttfamily ChemistryService, EvolutionService, NetworkService, AnalysisService} and {\small \ttfamily  UserManagementService}. 

One of the most compelling reasons to use Spring for service management is its comprehensive transaction support. Spring provides a consistent abstraction for transaction management that integrates very well with various data access abstractions. For remote access, the service interfaces can be easily injected by appropriate stubs. Spring supports for example Remote Method Invocation (RMI), Spring's HTTP invoker, JAX-RPC, JAX-WS or JMS.

Since the web client runs as a part of the application context, i.e., it lives inside the same server-side JVM (Java Virtual Machine) as Spring, all service calls are local. On the other side, the console client runs as a separate process and its calls are remote. More precisely, console clients requests are carried out by RESTFul Web Services and alternatively by JMS. In the future we might consider exposing a portion of services to 3rd parties, possibly other universities or teams, through REST.

\subsection{Cloud Computing}
\label{subsec:cloud_computing}


COEL's computational grid has been built on top of the GridGain In-Memory Computing Platform \cite{gridgain}. The GridGain HPC (High Performance Computing) library implements a scalable low-latency zero-deployment computational grid, which fits seamlessly into our Spring-backed IoC container (Section~\ref{subsec:ioc}).

COEL's grid currently consists of $19$ nodes with around $500$ cores. All nodes are hosted on Portland State University hardware, though the technology allows us to add any geographically remote resource, since the communication is carried out by TCP/IP protocol with optimized marshaling (serialization) of exchanged data. We plan to utilize existing grid technology to pool the resources with other geographically dispersed teams.

COEL's grid acts transparently, as a single computing resource. GridGain enables COEL's users to be more productive by eliminating the complexity of distributed computing. Regardless of a user's geographic location, they can add tasks to the grid from the COEL web page without much effort. When a user submits a task, after the chain of calls the request is ultimately received by the grid master node running within the application context. The task splits into many partial jobs, which are then distributed over the grid.

GridGain provides zero-deployment technology, so a new (slave) node could be added to the grid on-the-fly by registering with the master node identified by the IP address or domain name. Therefore the grid's topology might change freely during its lifetime. COEL's grid supports several enterprise features contributing to effective and robust execution of jobs. The grid keeps track of various node statistics such as CPU performance, execution time, and availability, which are constantly updated and utilized for adaptive job distribution such that high performing nodes obtain more jobs. Also, if a node disconnects from the grid, the exception is noted by a periodic heartbeat, and disconnected node's jobs are redistributed across the grid. Moreover, if a node finishes its execution sooner that expected and so it sits idle (its wait queue is empty), it steals jobs from other nodes. 

Due to the communication and task initialization overhead we execute only nontrivial tasks on the grid, with compute times that can last seconds, hours, or days. The main grid tasks include chemical ODE simulations, dynamics analyses, and evolutionary optimizations of rate constants.

\begin{figure}[htb!]
\centering
    \includegraphics[width=.9\textwidth]{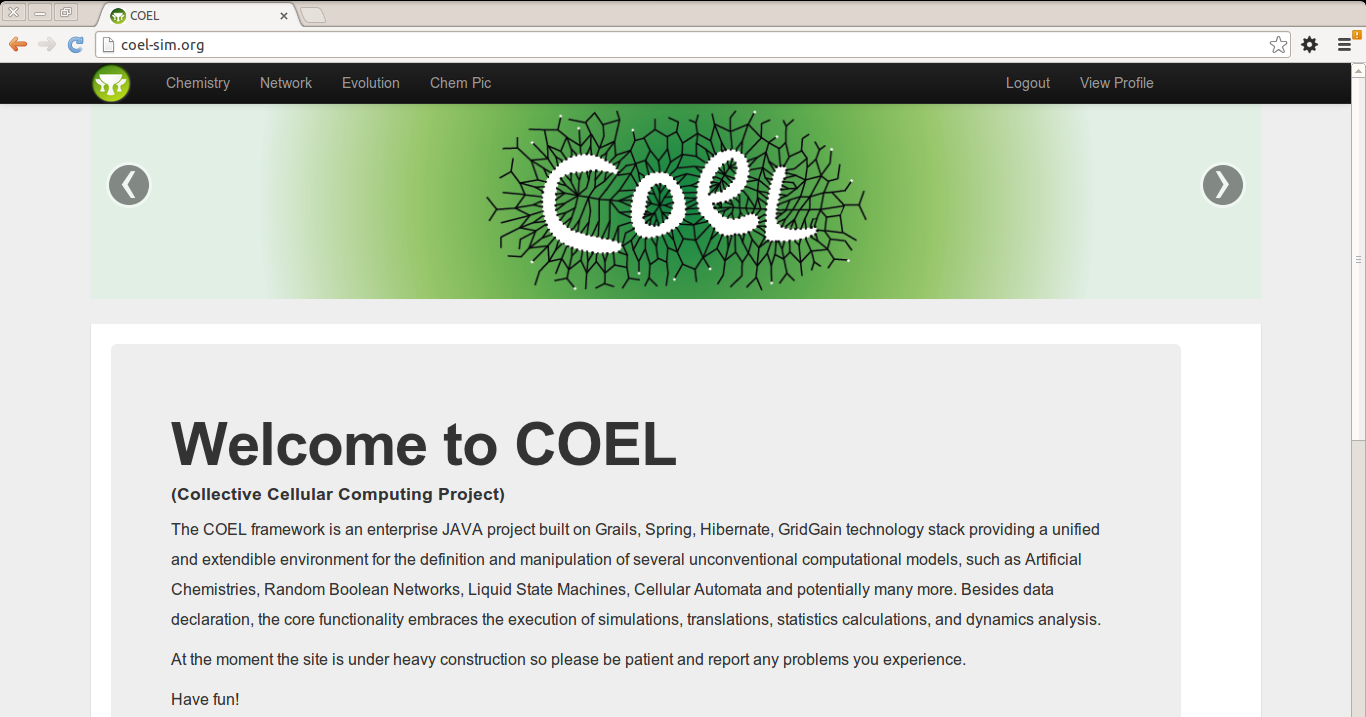}
\caption{COEL's home (welcome) page. URL: \href{http://coel-sim.org}{coel-sim.org}.}
\label{fig:coel-logo}
\end{figure}

\subsection{Web Client}
\label{subsec:grails}


COEL's web client is implemented in Grails \cite{grails}, which is a powerful web 2.0 framework using the Groovy dynamic language for the Java Virtual Machine. JVM compatibility means that Java, Groovy, and Scala source compiles into Java byte code, hence these three languages are natively inter-callable. Grails follows the "Convention over Configuration" approach, which emphasizes standard (conventional) naming, binding and data flow, so the structure of the application is simply implied if it is not explicitly configured. This approach is heavily utilized in a function called scaffolding, which based on a domain object structure generates dynamically at runtime the controller with associated web pages, providing basic CRUD operations without any effort. As a matter of fact, we could build a COEL prototype web client just with a few lines of code. Grails internally uses Spring IoC for dependency injection and bean creation. Furthermore, Grails was officially incorporated into Spring at the end of 2008.

The web front-end relies heavily on Javascript provided by the jQuery library~\cite{jquery}, which makes UI interactive and intuitive and moves a part of data processing and visualization directly to the web browser. For instance, although COEL runs all simulations server-side, if a user wishes to see a chart, e.g., of species concentration traces, COEL sends the user raw data which is transformed into a chart by client-side Javascript using Google Charts API. For styling and some widgets we used the Bootstrap library~\cite{bootstrap} created by Twitter. 


\subsection{Persistence}
\label{subsec:persistence}



The persistence layer consists of DAOs (Data-Access Objects) wrapping storing, retrieving, deleting, and filtering functionality for domain objects. To map an object-oriented domain model to a traditional relational database we use Hibernate \cite{hibernate}, an object-relational mapping (ORM) library for the Java language. DAOs and Hibernate are widely supported by Spring, which offers hooks for fast integration.

Hibernate solves Object-Relational impedance mismatch by replacing direct persis-tence-related database accesses with high-level object handling functions. Hibernate provides declarative strategy for persisting data. We define a mapping of columns, reference metadata and inheritance strategy mapping. Hibernate handles details about persistence implementation, like SQL statements and JDBC connection creation. To obtain data we use SQL or the Hibernate query language (HQL). The actual translation from the POJO to JDBC result set is automatic. Hibernate also uses various optimization strategies, such as cache and DB access optimization.

We believe that it is imperative to store data in a structured database, enabling prompt retrieval, searching and post-processing. PostgreSQL \cite{postgresql} is a mature open source database providing standard SQL/PLSQL language support with numerous additional features. The decision to select PostgreSQL as DB provider was driven mainly by the following factors: a lot of hands-on experience, a comprehensive console as well graphical UI (PgAdmin), an open source license, and support for array data types, useful for storing scientific vector data. The database model currently contains about 90  tables. To assure compatibility for each version of COEL we migrate data by a set of SQL scripts. Also, each day the whole database is dumped (backed-up), so we could restore the state of the DB to a certain date and time very quickly. That means our data is stored safely in structured and indexed format.

\subsection{Build, Deploy, and Testing}
\label{subsec:build-deploy-testing}
\begin{figure}[htb!]
  \centering
  \includegraphics[width=.5\textwidth]{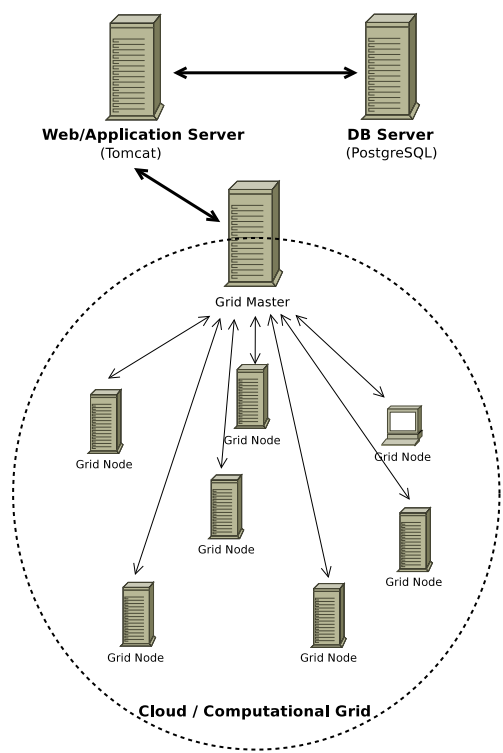}
  \caption{Diagram showing a physical deployment of COEL's components.}
  \label{fig:deploy}
\end{figure}

To build COEL's project and to maintain its library dependencies, we use Apache Maven~\cite{maven}. For a new application version we run a set of JUnit tests, which guarantee that the core functionality works as expected. After that, COEL is deployed to the Tomcat application server. Figure~\ref{fig:deploy} shows a deployment schematic of COEL's components over several resources (machines), each running some part of the application: the database server, the application server, and the cloud. Due to the extendability of the computational cloud, the number of resources is not bounded. Also, note that the database server and the application server are currently hosted on the same machine.

COEL currently has about 30 users (exclusively from the NSF project ``Computing with Biomolecules'' and Portland State University), 5 of which are active, i.e., they access COEL on a daily basis. Once COEL will be available to the research community we expect the number of users to grow to hundreds, which would require more resources and more rigorous testing. If the users find a production issue or want to recommend a new feature, they will be able to submit a report through a Jira issue tracking system. More than $60$ issues and new feature requests have be reported so far internally. Currently, the development of COEL is largely driven by the authors' research needs.

%% file: 4_conclusion.tex
\section{Conclusion and Future Work}
\label{sec:conclusion}

In this paper, we presented a new web-based chemistry simulation framework, COEL. Its modern layered architecture includes a scalable computational grid, a user-friendly and interactive web UI, and the safe and transactive persistence of chemistry models and simulation results. Its wide range of features primarily target chemistry simulations, GA optimization of rate constants, performance evaluations, and dynamics analysis. We paid particular attention to general usability and lightweight and fluid layout, and embedded data visualization using Google's charting engine.

COEL can be used without any installation, and from any web browser. As such, it is easier to start using and has a larger potential audience than existing desktop-application based frameworks. Keeping COEL in the cloud allows for easy collaboration and sharing of results, and makes it simple to build upon another's work.


COEL's computational grid utilizes CPU resources only, however, it would be beneficial to extend the grid over GPUs as well. GridGain, our current computational grid library, does not provide native support for GPUs. On the other hand, we argue that reimplementing all tasks and business logic in (J)CUDA or OpenCL and maintaining two code branches would not be feasible. Therefore, we plan to explore transparent compilation mechanism such as Aparapi, where a single Java code compiles to CPU and GPU version transparently and gets executed based on resource availability.

Furthermore, we often face the situations when we want a newly submitted task to be executed as soon as possible, or we want to associate more CPU time to the tasks of a certain user. To achieve that we would like to assign priorities to the tasks based on their type and users' privileges.

As mentioned in Section~\ref{subsec:service} we might consider exposing certain services and routines through RestFul API so 3rd party applications could call, integrate and tailor COEL's functionality for their needs.

To improve the quality of chemistry ODE-based simulations we plan to integrate the standard LSODA solver. Also, to provide an alternative to the deterministic ODE solvers our goal is to introduce a stochastic simulator based on the Gillespie method \cite{gillespie1977}. The Gillespie method simulates each reaction step stochastically on a molecular level \cite{turner04,jahn10}. It is computationally more demanding than ODE integration, however, it is physically more realistic, especially if the number of molecules in the system is low. Therefore, COEL is currently best-suited to simulate systems with large numbers of each chemical species.

Also, we plan to introduce more advanced sharing permissions, so each user could specify with which group or user he wants share the models and results for viewing and editing.

Last but not least, our vision for COEL is to become a common platform for diverse unconventional computing models. One step toward that goal is a new Network module, which will simulate complex spatial, random, or layered networks with configurable node functions and interaction series.